# Prefect spin filtering and spin inverting using a MoS$_2$ quantum dot

Hamidreza Simchi[*,1,2], Mahdi Esmaeilzadeh[**,1], Hossein Mazidabadi[1], and Mohammad Norouzi[1]

[1]Department of Physics, Iran University of Science and Technology, Narmak, Tehran 16844, Iran

[2]Semiconductor Technology Center, Tehran Iran

We consider a quantum dot of MoS$_2$ which is made by decreasing the width of finite length of a MoS$_2$ zigzag nanoribbon. The spin-dependent conductance and spin-polarization of the device are studied in presence (absent) of an external electric field perpendicular to the molybdenum plane by using the tight-binding non-equilibrium Green's function method. It is shown that in absence of the electric field, the deformation of ribbon structure causes spin-splitting in the dot and non-prefect spin filtering is seen. Therefore, the technique could be used for designing spin-dependent devices of MoS$_2$. Also, we show that the device could behave as a prefect spin filter and spin inverter under applying an external electric.

I. **Introduction**

For manufacturing diodes and transistors, we need semiconductor materials. Semiconductor materials have a finite band gap.[1] For instance, armchair garphene nanoribbons have a finite gap[2] and also, bilayer graphene under perpendicular electric field has a gap.[3] Nowadays, scientists and technologist try to introduce an atomic monolayer bulk sample which has a finite gap. Silicene is one of the promising candidates for post garphene materials.[4] Another promising candidate is transition metal dichalcognides (TMDCs) with formula $MX_2$, where M is a transition metal and X is a chalcogen.[5,6] Molybdenum disulfide ($MoS_2$) is a TMDC semiconductor with an indirect band gap.[7,8] It has been shown that, inversion symmetry breaking together with spin-orbit coupling leads to coupled spin and valley degree of freedoms in monolayer of $MoS_2$ and other group-VI dichalcogenides.[9] It means that, it is possible to control both spin and valley in these two dimensional (2D) materials.[9] The effect of strain on band gap (direct or indirect), binding energy and optical gap of $MoS_2$ have been studied.[10,11] It has been shown that zigzag (armchair) nanoribbons of $MoS_2$ show the ferromagnetic (nonmagnetic) and metallic (semiconductor) behavior.[12] Also, armchair become semiconducting when passivating with hydrogen.[13] Zigzag nanoribbons of $MoS_2$ exhibit unusual magnetic properties regardless of passivation.[13] Yue et al. have shown that the band gap of monolayer armchair $MoS_2$ nanoribbons can be significantly reduced and be closed by applying a transverse electric field, whereas the band gap modulation is absent under perpendicular electric field.[14] Klinovaja et al. have studied spintronics in $MoS_2$ monolayer quantum wire.[15] They have shown that, an external electric field can interplay between intrinsic and Rashba spin orbit interaction and in consequence spin filtering and prefect spin polarization could be seen for special conditions.[15] Rostami et al. have studied the effect of a perpendicular magnetic field on the electronic structure and charge transport of a monolayer $MoS_2$ nanoribbon.[16] They have shown that, the transport channels are chiral for one of the spin component in a low-hole doped case.[16] We have shown that, one can tune spin transport using a $MoS_2$ nanoribbon if he/she uses both transverse and perpendicular electric fields, simultaneously.[17]

In this paper, we consider a quantum dot of $MoS_2$ with zigzag edge. It is made by decreasing the width of finite length of a $MoS_2$ zigzag nanoribbon. The spin transport property and spin polarization of the device are studied in presence (absent) of an external electric field perpendicular to Mo-plane, using tight-binding non-equilibrium Green's function method. We show that, the deformation in ribbon structure causes spin-splitting. Therefore, the technique can be used for designing spin-dependent devices of $MoS_2$. Also, it is shown that, the quantum dot could behave as a perfect spin filter and spin inverter under applying the external electric field. The structure of article is as follow. In section II, the calculation method is presented. Results and discussion are presented in section III and in section IV the summary and conclusion are presented.

II. **Calculation method**

Different tight binding Hamiltonians have been proposed for monolayer of $MoS_2$ before.[18-22] Here, we use Hamiltonian proposed in Ref.20 by considering $4d-orbitals$ of the Mo-atom, $d_{x^2-y^2}, d_{xy}$ and $d_{z^2}$, and two $3p-orbitals$ of S-atom, $p_x, p_y$. The Hamiltonian $H$ and overlap $S$ matrices can be written as:[20]

$$H(k) = \begin{pmatrix} H_a & H_t & H_t \\ H_t^\dagger & H_b & 0 \\ H_t^\dagger & 0 & H_{b'} \end{pmatrix} \quad (1)$$

$$S = \begin{pmatrix} 1 & S & S \\ S^\dagger & 1 & 0 \\ S^\dagger & 0 & 1 \end{pmatrix} \quad (2)$$

It should be noted that $H_a$ is on-site energy Hamiltonian of Mo-atom and $H_b$ and $H_{b'}$ are on-site energy Hamiltonian of S-atom above and under the Mo-plane, respectively.[20]

$$H_t = \begin{pmatrix} t_{11}f(k,\omega) & -e^{-i\omega}t_{11}f(k,-\omega) \\ t_{21}f(k,-\omega) & t_{22}f(k,0) \\ -t_{22}f(k,0) & -e^{-i\omega}t_{21}f(k,\omega) \end{pmatrix} \quad (3)$$

where $t_{11}$ is hopping integral between orbital $d_{z^2}$ of Mo-atom and orbital $p_x + ip_y, p_x - ip_y$ of S-atom. Also, $t_{21}$ is hopping integral between orbital $d_{x^2-y^2} + id_{xy}$ ($d_{x^2-y^2} - id_{xy}$) of Mo-atom and orbital $p_x + ip_y$ ($p_x - ip_y$) of S-atom. Finally, $t_{22}$ is hopping integral between orbital $d_{x^2-y^2} + id_{xy}$ ($d_{x^2-y^2} - id_{xy}$) of Mo-atom and orbital $p_x - ip_y$ ($p_x + ip_y$) of S-atom.

$f(k,\omega) = e^{i\vec{k}\cdot\vec{\delta_1}} + e^{i(\vec{k}\cdot\vec{\delta_2}+\omega)} + e^{i(\vec{k}\cdot\vec{\delta_3}-\omega)}$ is the structure factor with $\omega = 2\pi/3$, in plane momentum $\vec{k} = (k_x, k_y)$, and in-plane components of the lattice vector $\vec{\delta}_{i\pm}$ ($i = 1,2,3$).[20] The overlap matrix $S$ is defined similar to $H_t$ but with hopping integral $s_{\mu\nu} = 0.1 t_{\mu\nu}$.[20] The spin-orbit coupling term (SOCT) $H_{SO}^{Mo} = \lambda\, diag(0, s, -s)$, where $\lambda = 0.08$ eV is the spin-orbit coupling constant and $s = \pm 1$.[20] It should be noted that the most important contribution of Mo atoms is considered in SOCT.[20] The values of on-site energies and hopping integrals and also vectors $\vec{\delta}_{i\pm}$ have been given in Ref.20 and thus we do provide them in the present study. The spin transport properties of the elongated MoS$_2$ device is studied using a divide and conquer tight binding approach.[23] By using the Eqs. (1) to (3), we can write the tight binding Hamiltonian of MoS$_2$ quantum dot (called $H_C$). The Green's function of the dot can be calculated by using[24-26]

$$\xi^{\uparrow(\downarrow)} = ((E+i\eta) \times I - H_C^{\uparrow(\downarrow)} - \sum_L - \sum_R)^{-1} \qquad (4)$$

where $E$ is the electron energy, $\eta$ is an infinitesimal number, $I$ is the unit matrix, and $\uparrow(\downarrow)$ stands for spin up (down). The self energies $\sum_{L(R)}$ and the coupling matrices $\Gamma_{L(R)}$ can be calculated by using the surface Green's function.[24-26] The surface Green's function of left ($g_L$) and right ($g_R$) leads could be calculated by suing the Sancho's algorithm.[27] Using Eq.(4), the spin dependent conductance is given by[24-26]

$$G^{\uparrow} = \mathrm{Im}[Trace(\Gamma_L \times \xi^{\uparrow} \times \Gamma_R \times \xi^{\uparrow\dagger})],$$
$$G^{\downarrow} = \mathrm{Im}[Trace(\Gamma_L \times \xi^{\downarrow} \times \Gamma_R \times \xi^{\downarrow\dagger})], \qquad (5)$$

where, Im denotes the imaginary part. The spin-polarization of transmitted electrons is defined by[24-26]

$$P_S \equiv \frac{(G^{\uparrow} - G^{\downarrow})}{(G^{\uparrow} + G^{\downarrow})}. \qquad (6)$$

In the absence of the external electric field, there is symmetry between up and down sulfur atoms and the on-site energy of sulfur atom is equal to $\varepsilon_b = \varepsilon_{b'} = 5.53$ eV.[20] When the external electric field is applied $\varepsilon_b + E_z a/2$ and $\varepsilon_{b'} - E_z a/2$ are on-site energy of S-atom above and under the Mo-plane, respectively. Since, $\varepsilon_b = \varepsilon_{b'}$ the potential difference between to sulfur atoms is equal to $E_z a$. It is noted that $E_z$ is the strength of electric field in eV A$^{0-1}$ and $a = 3.116$ A$^0$ is the distance between up and down sulfur atoms.

### III.   Results and discussion

Let us to consider a monolayer of zigzag nanoribbonn of MoS$_2$. The nanoribbon is made by repeating a super cell which includes 8 Mo-atoms and 16 S-atoms (totally 24 atoms). Then we assume that the nanoribbon is deformed by using some kinds of lithography process such that at finite length of ribbon, its wide is decreased by 4 Mo-atoms and 8-S-atoms. Castle et al. have explored and developed a simple set of rules that apply to cutting, pasting, and folding honeycomb lattices.[28] The cut region, which is called MoS$_2$ quantum dot, is connected to the two semi-infinite nanoribbon of MoS$_2$. Fig. 1 shows the schematic of deformed monolayer of zigzag nanoribbon of MoS$_2$ . The Mo-atoms are shown in green color while the sulfur atoms are shown in yellow color.  Fig. 2(a) shows seven bands structure of MoS$_2$ which is found by using Eqs. (1) to (3).[17] As it shows, the two conduction bands are spin splitting and the energy band gap, $E_g = 5.34$ eV at $k_x = 0.0$. Fig. 2(b) shows conductance versus electron energy of zigzag nanoribbon of MoS$_2$ for $k_x = 0.0$ . As it shows, no spin dependent is seen and transmission gap, $E_T = 5.60$ eV, which is at order $E_g$ .Fig. 3 shows conductance versus electron energy of quantum dot, when $E_z = 0.0$ VA$^{0-1}$.[17] Hod et al. have shown that there is a critical length such that the conductance of an elongated graphene nanoribbon can be equal to the conductance of an infinite nanoribbon.[29] For lengths smaller than the critical length, the conductance curve is localized around some specific electron energies.[29] By comparing Fig.2 (b) and Fig. 3, it can be concluded that the conductance curve of MoS$_2$ quantum dot is localized around some specific electron energy and is spin-dependent. Of course, the conductance decreases due to the decrement of the number of transport channels through quantum dot.[29] Before lithography, 10 Mo-atoms are placed on the top-edge and 10 S-atoms are placed on the down-edge of the quantum dot region i.e., only one type of atoms exist on each edge. After lithography, quantum dot is composed by 48 Mo-atoms and 96 ($48 \times 2$) S-atoms (totally, 144 atoms). But among them, 8 Mo-atoms and 4 S-atoms are placed on the top-edge and 8 S-atoms and 4 Mo-atoms are placed on the down edge of the quantum dot (see Fig. 1). Therefore, on both edges there are not only Mo-atoms but also S-atoms. We think it causes the spin splitting in quantum dot. Therefore, the technique can be used for designing spin-dependent devices of MoS$_2$.

Fig. 4 shows spin polarization $P_S$ versus electron energy of quantum dot, when $E_z = 0.0$ VA$^{0-1}$ As the figure shows, the spin polarization is near to $\pm 1$ for some values of electron energies. It should be noted that in transmission gap region we ignore $P_S$ since conductance is very small in the region.

It is desirable if we can tune spin polarization by applying an external electric field. We assume that an external electric field $E_z = -2, -1, 0.0, 1, and, 2$ eV nm$^{-1}$ is applied perpendicular to the Mo-plane. Fig.5 (a) shows spin polarization $P_S$ versus electron energy when $E_z$ is present. As the figure shows, $P_S$ varies by variation of $E_z$. As Fig. 5(b) shows in absence of electric field, spin polarization is at range $-0.4 \leq P_S \leq 0.4$ (curve in green color). When electron energy is at range $2.6 \leq E \leq 2.8$ eV and $E_z = 1,2(-1,-2)$ eV nm$^{-1}$ the spin polarization is equal to $P_S \simeq -1(+1)$. Similarly, when electron energy is at range $2.9 \leq E \leq 3.2$ eV and $E_z = 1,2(-1,-2)$ eV nm$^{-1}$ the spin polarization is equal to $P_S \simeq +1(-1)$. Therefore, the MoS$_2$ quantum dot behaves not only as a prefect spin filter but also as a spin inverter. Also as Fig. 5(c) shows, for some specific values of electron energy when $-3.4 \leq E \leq -3$ eV the device behaves as a prefect spin filter. It should note that as Fig. 6 shows the conductance of quantum dot is not small for some values of electron energy when the device behaves as prefect spin filter or spin inverter. Therefore, we could save the spin polarization for these values of energies. For better demonstrating, Fig. 7(a), (b) and (c) show three dimensional graph of spin polarization, conductance of spin up, and conductance of spin down electrons versus electron energy and external electric field $E_z$ for, respectively. As they show the spin polarization can be only equal to ±1 for specific values of electron energy.

IV. **Summary and conclusion**

It has been shown that, the conductance of monolayer zigzag MoS$_2$ nanoribbon is no spin-dependent at $k_x = 0.0$. Then we have assumed that, the nanoribbon is deformed and its wide is decreased by 4 Mo-atoms and 8-S-atoms. The new device is called quantum dot. We have shown that, the conductance of quantum dot is spin-dependent and therefore, the technique could be used for designing the spin-dependent devices of MoS$_2$. Also, it has been shown that, the quantum dot behaves as a perfect spin up filter and spin inverter when an external electric field is applied perpendicular to Mo-plane. In the other words, one can tune the spin filtering and spin inverting in MoS$_2$ quantum dot by applying the electric filed.

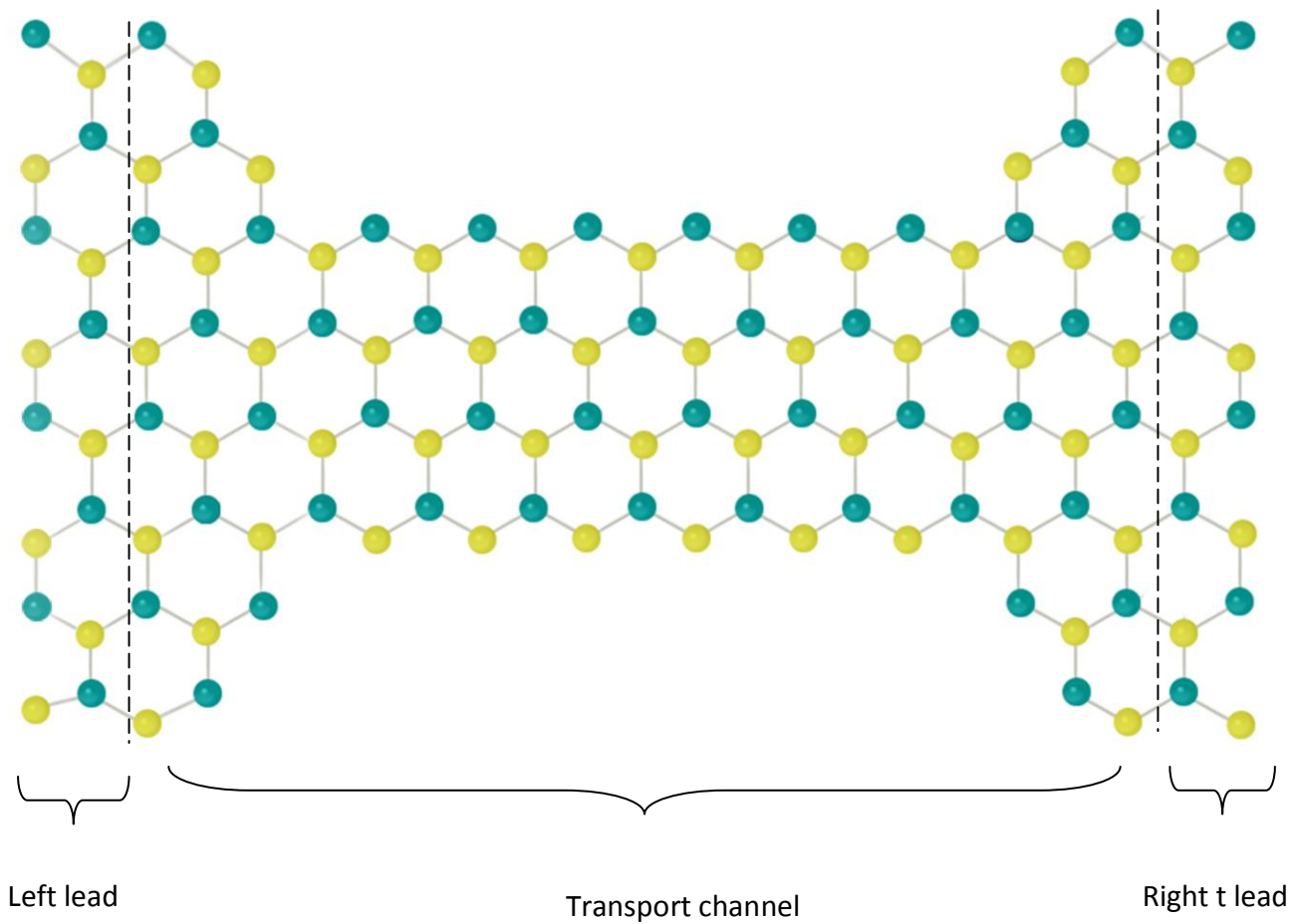

Fig. 1 A MoS$_2$ quantum dot connected to two semi-infinite MoS$_2$ nanoribbons from left and right. Mo -atom and S-atom is shown in green and yellow color, respectively.

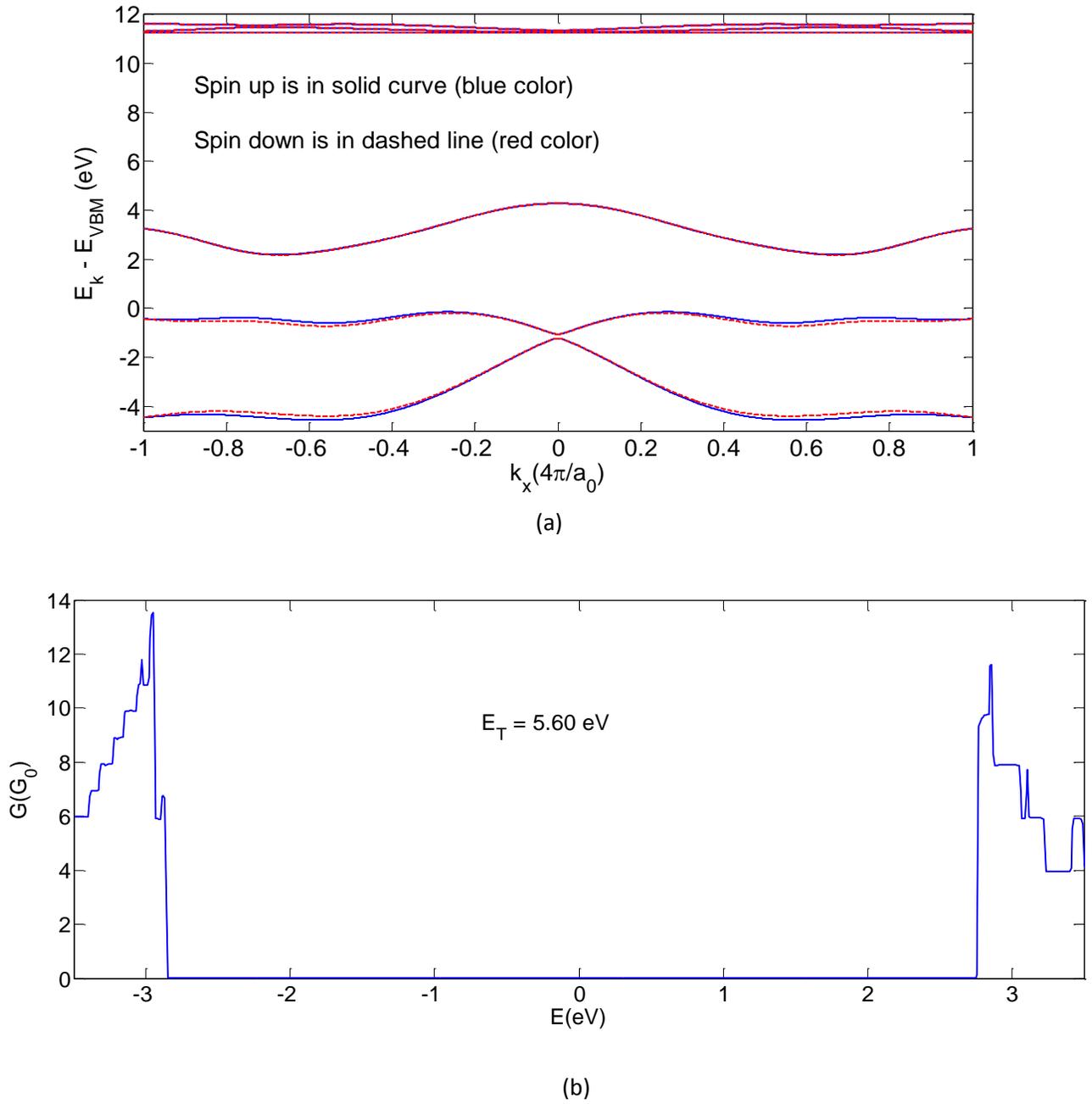

Fig. 2 (Color online) (a) Band structure of monolayer of MoS$_2$ consisting of seven bands in which two are spin split in the valence band. The valence band minimum energy is $E_{VBM} = -5.73$ eV.[17] (b) Conductance versus electron energy of zigzag nanoribbon of MoS$_2$. Here, $E_z = 0.0$ V nm$^{-1}$.

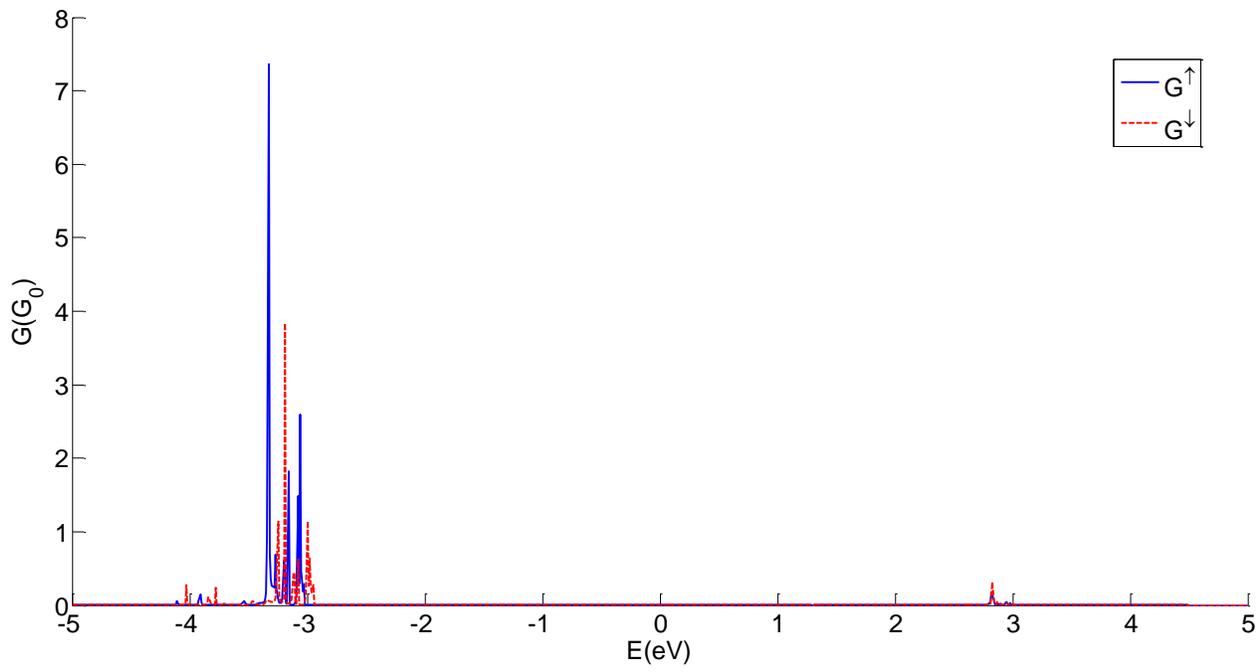

(a)

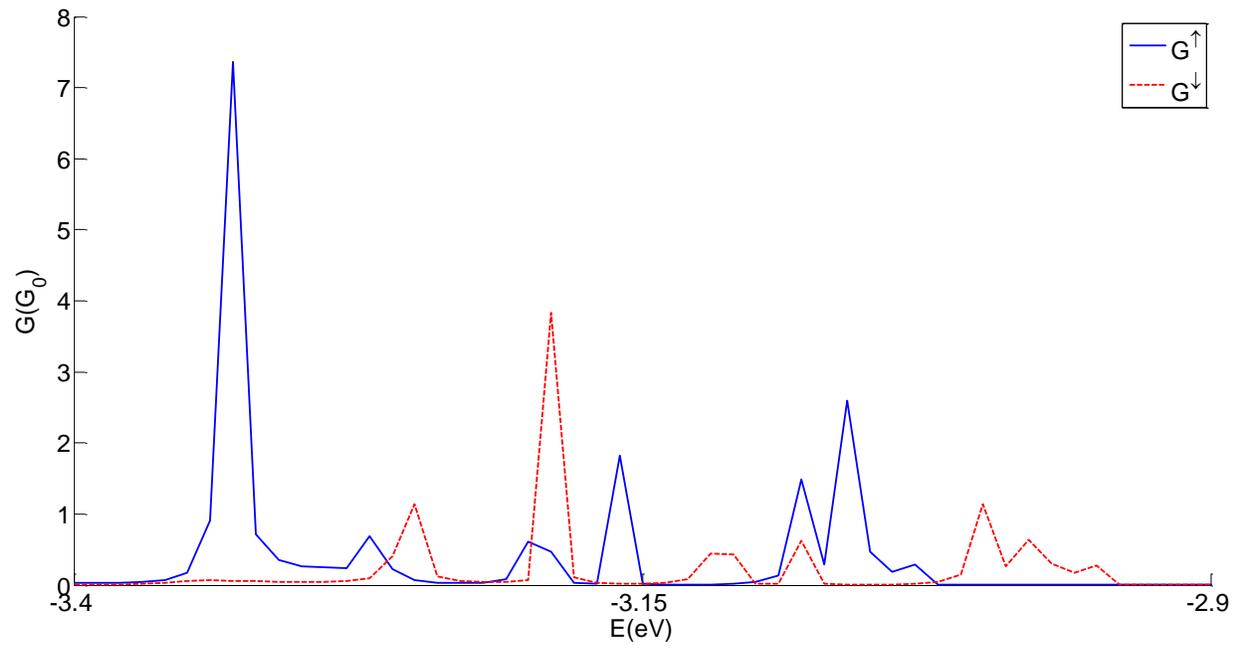

(b)

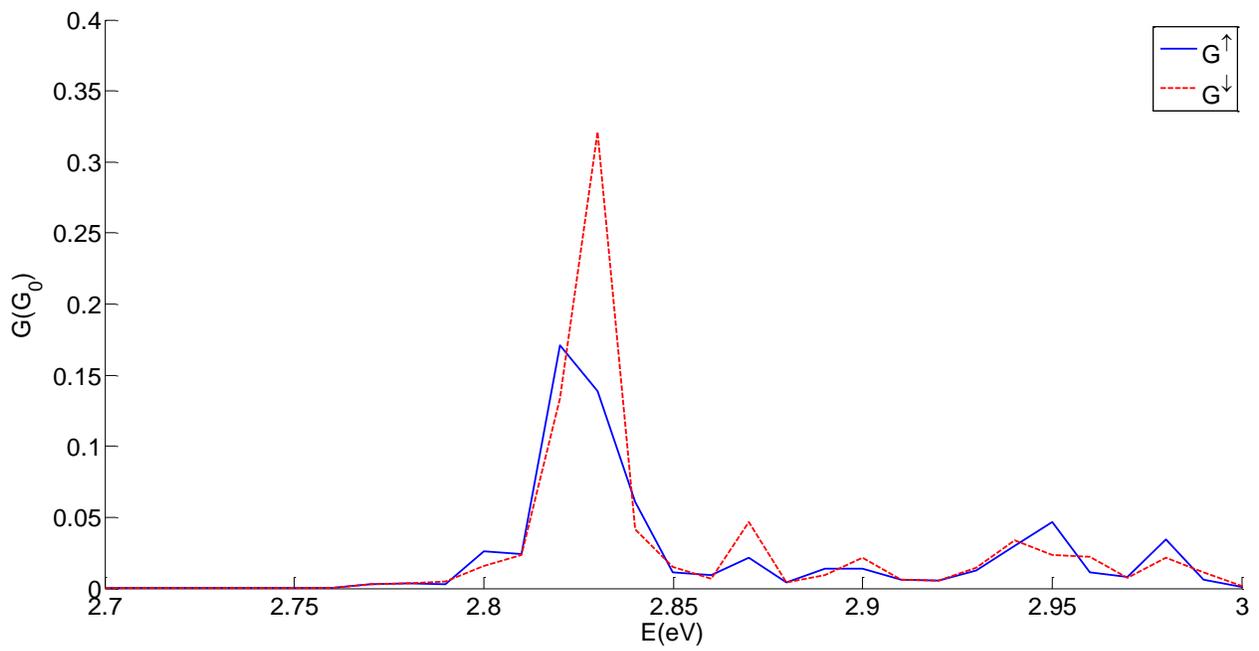

(c)

Fig. 3 (Color online) Conductance versus electron energy of quantum dot of MoS$_2$ (a) for $-5 \leq E \leq 5$ eV, (b) $-3.4 \leq E \leq 2.9$ eV, and (c) $2.7 \leq E \leq 3$ eV. Here, $E_z = 0.0$ V nm$^{-1}$.

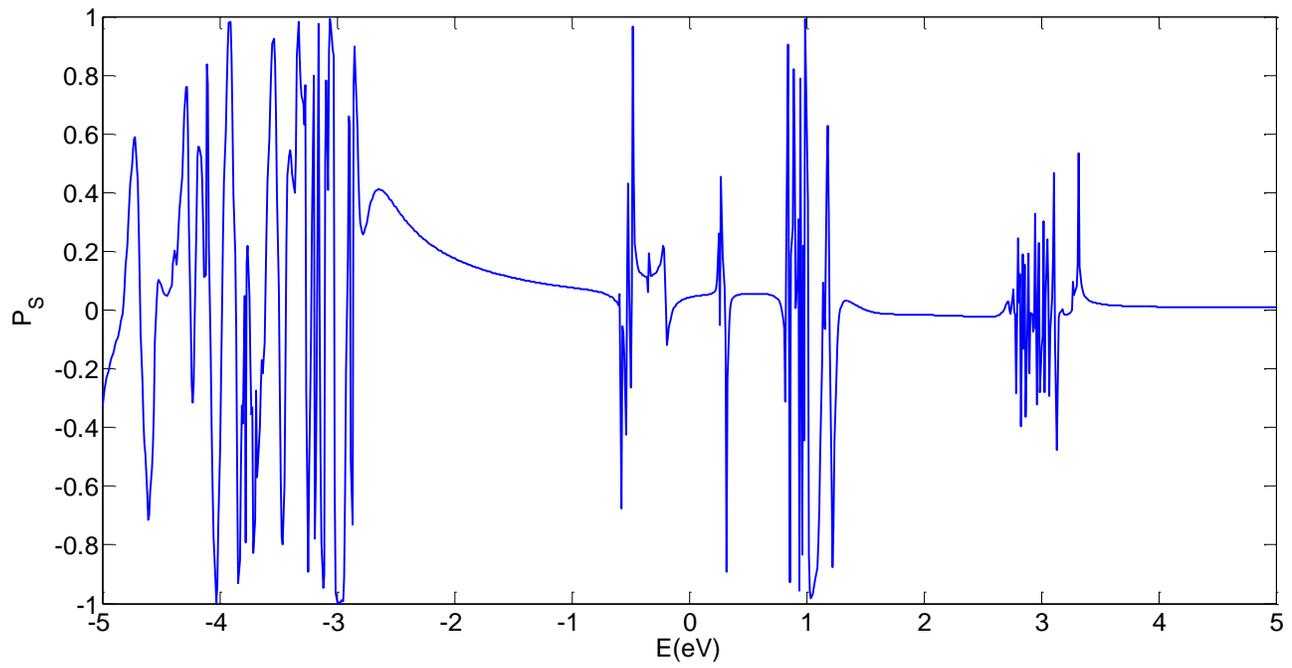

Fig.4 Spin polarization versus electron energy of quantum dot of MoS$_2$. Here, $E_z = 0.0$ eV nm$^{-1}$.

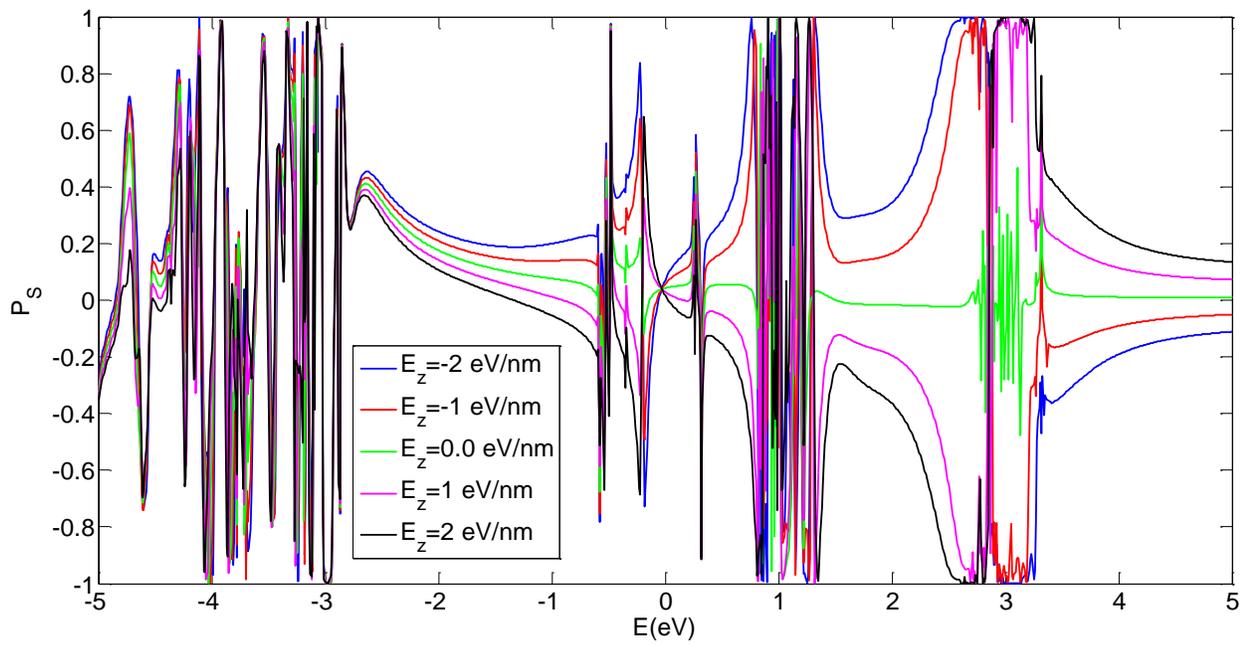

(a)

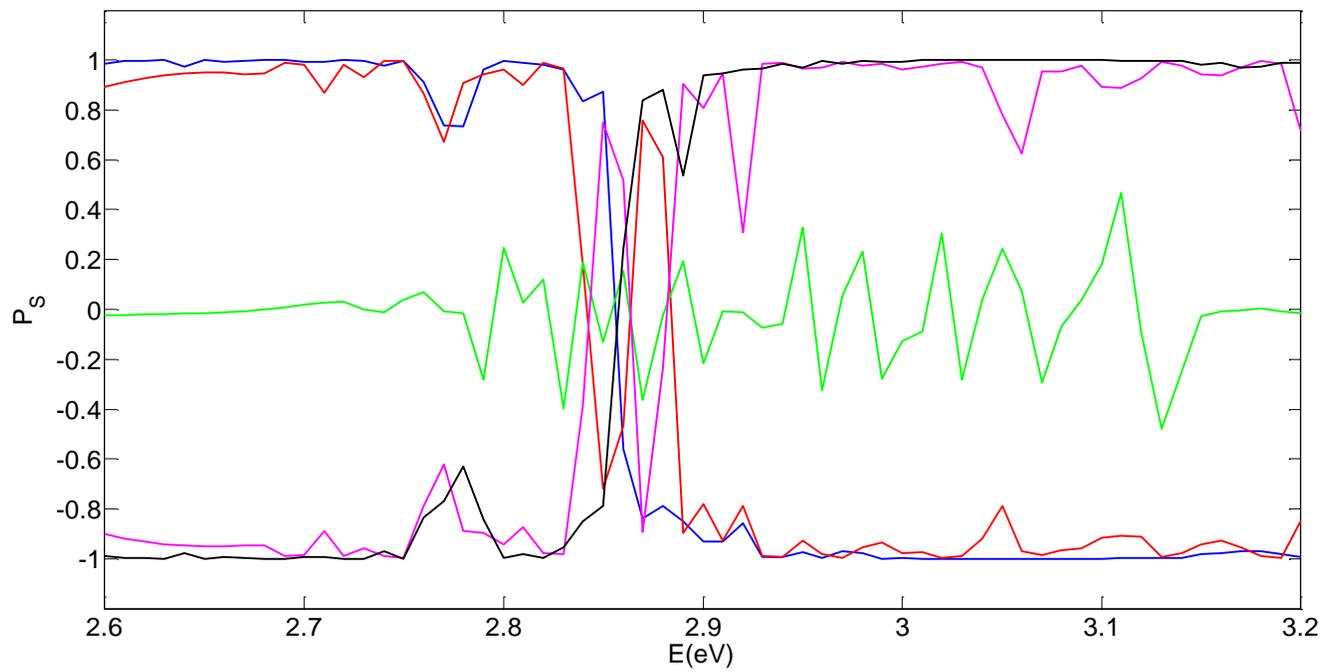

(b)

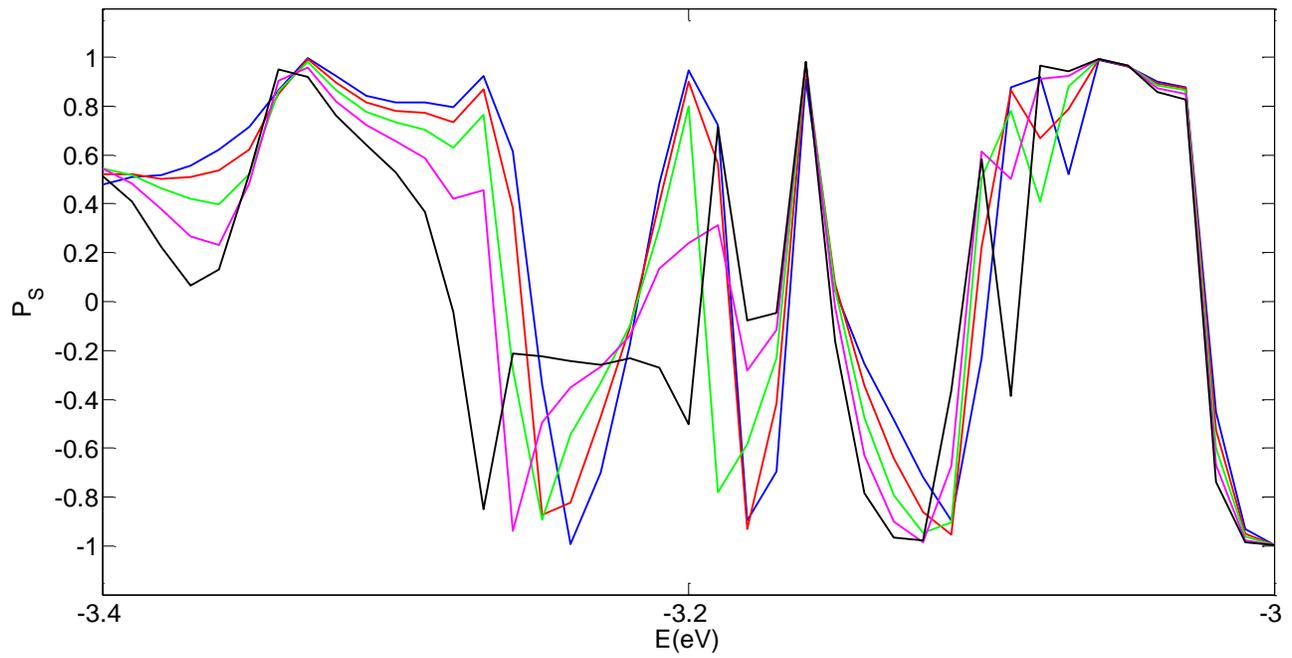

(c)

Fig. 5 (Color online) Spin polarization versus electron energy, when $E_z$ is present, (a) for $-5 \leq E \leq 5$ eV, (b) for $2.6 \leq E \leq 3.2$ eV, and (c) for $-3.4 \leq E \leq -3$ eV.

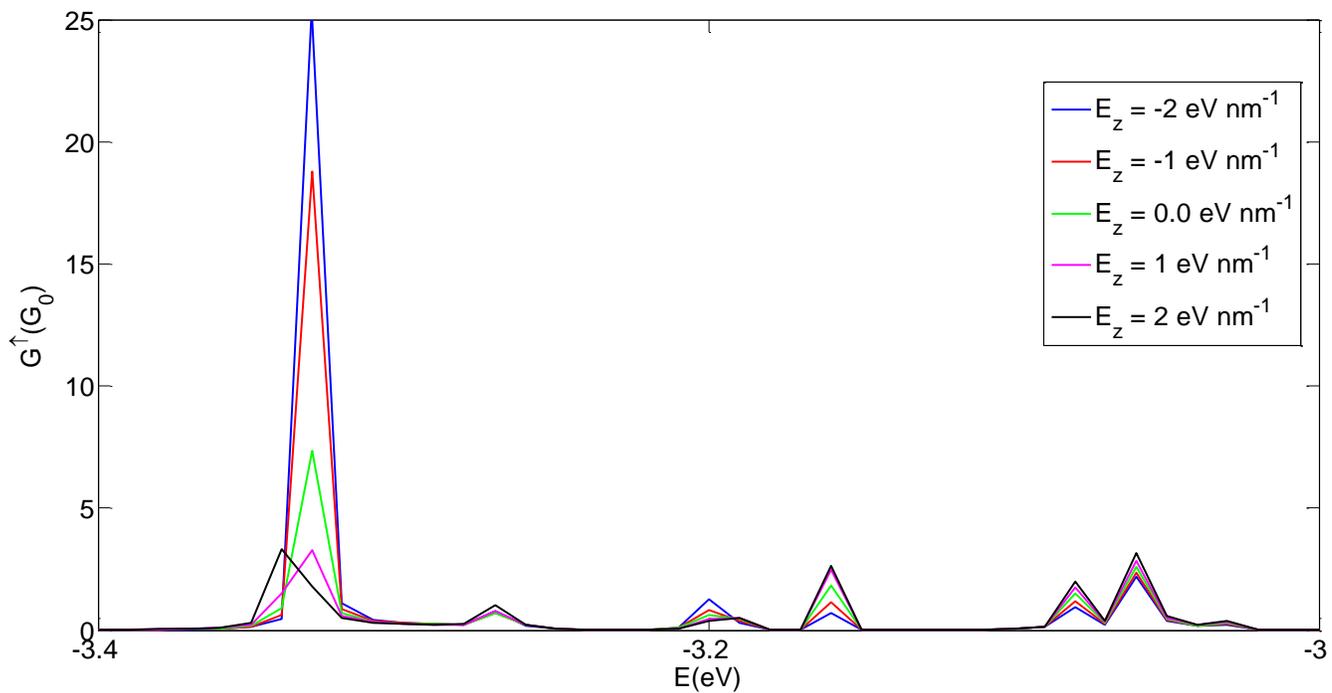

(a)

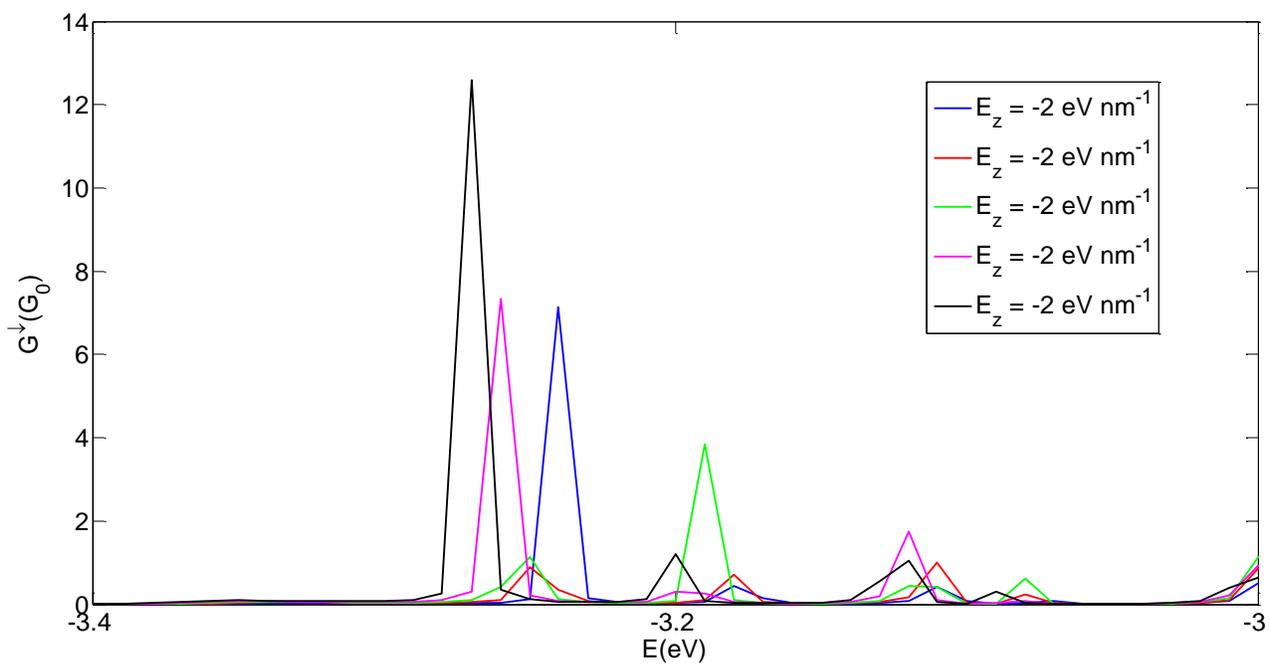

(b)

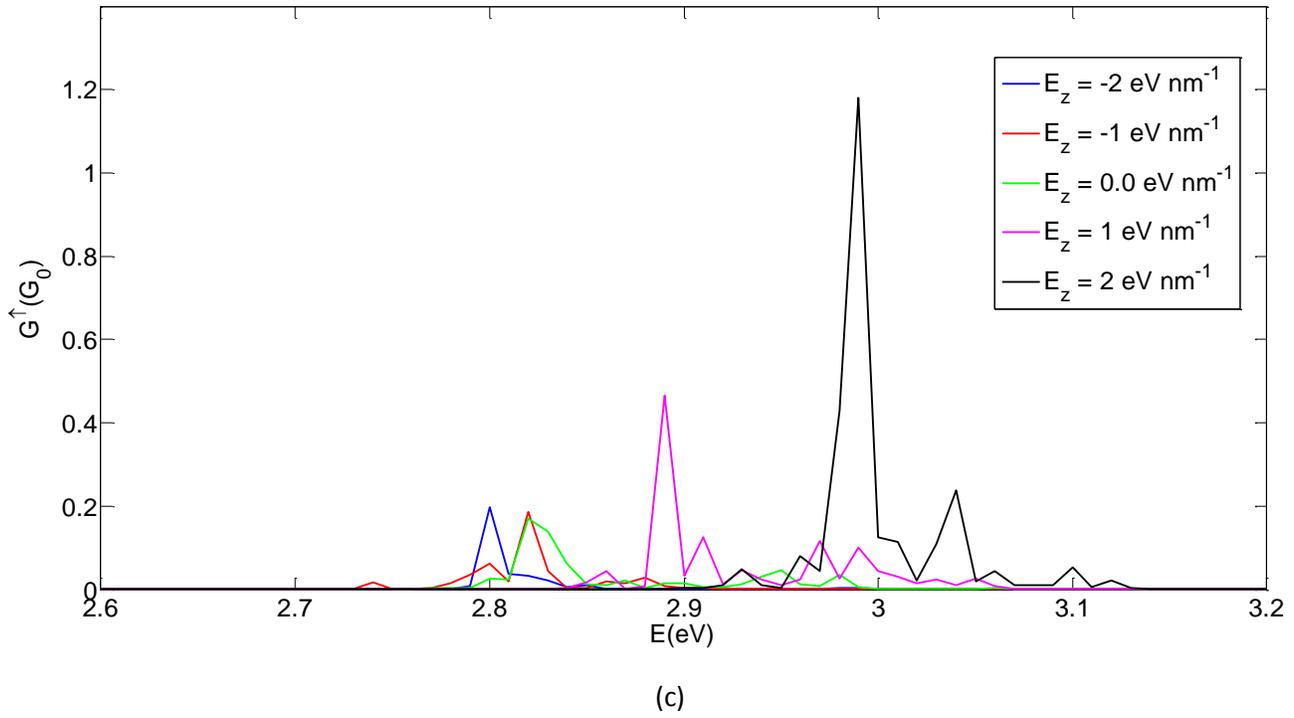

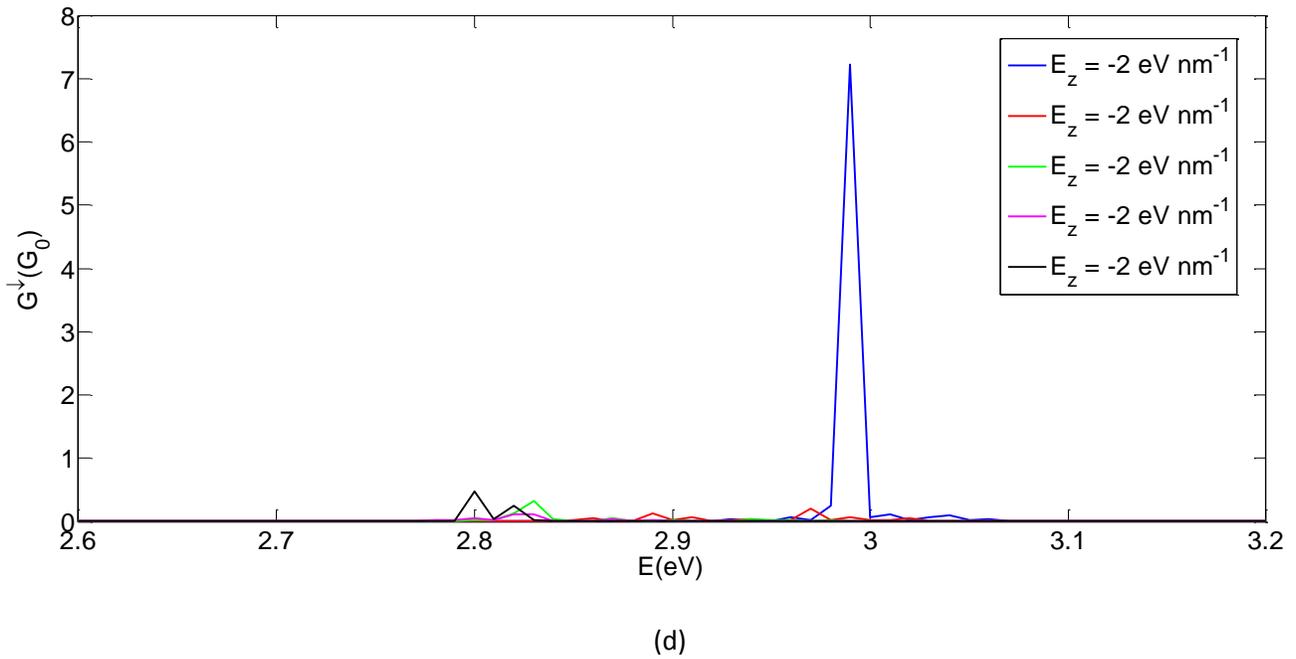

Fig. 6 (Color online) Conductance versus electron energy of quantum dot of MoS2 (a) and (c) for spin up and (b) and (d) for spin down electrons when $E_z$ is present.

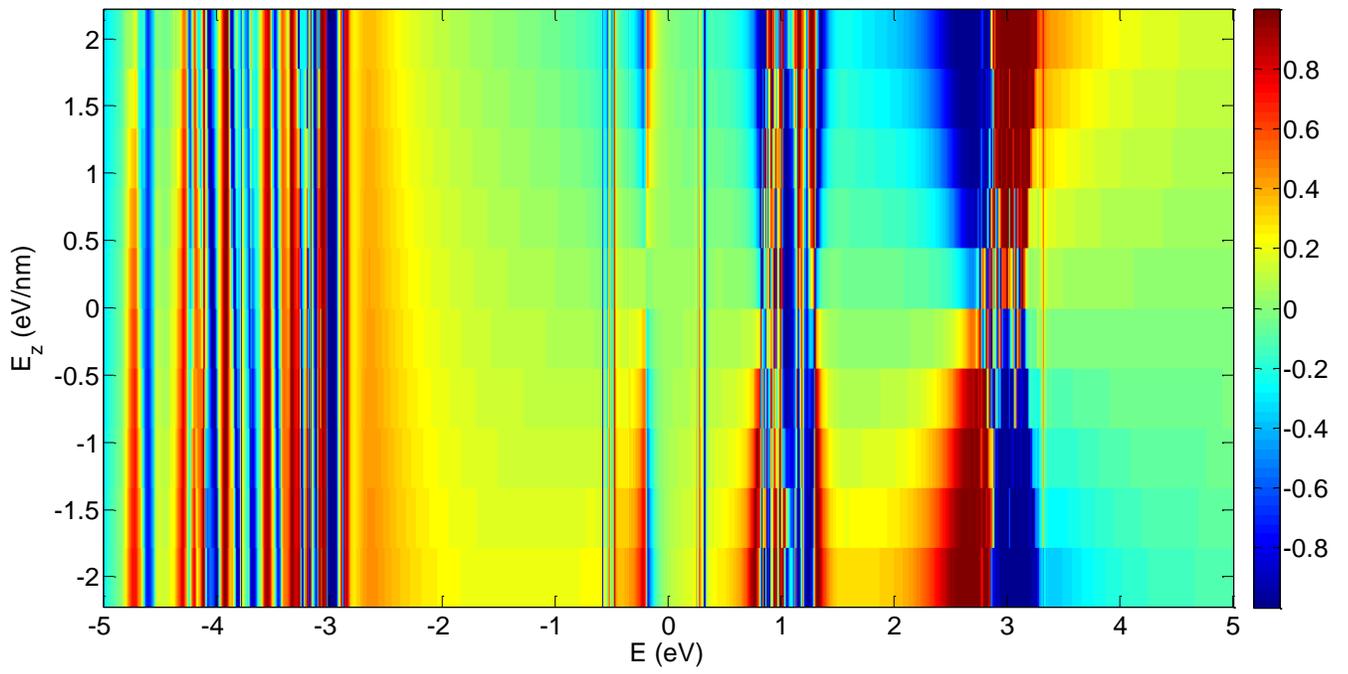

(a)

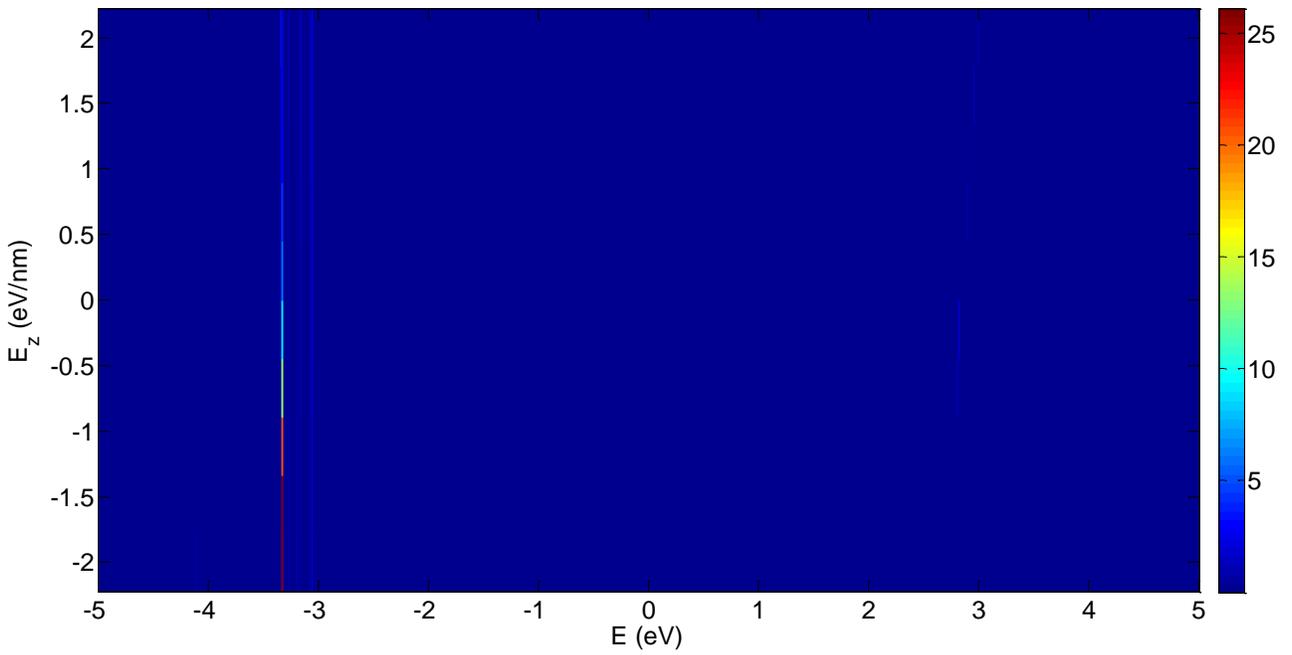

(b)

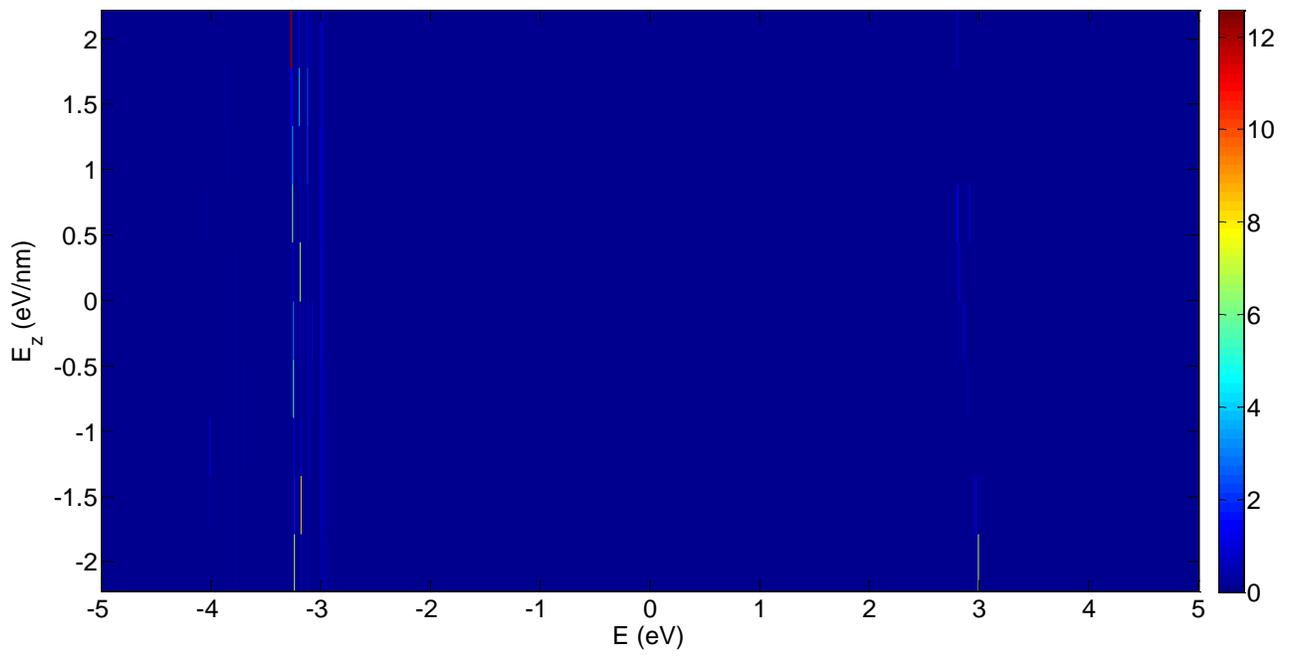

(c)

Fig. 7(Color online) (a) spin polarization, (b) conductance of spin up $G^{\uparrow}$, and (c) conductance of spin down $G^{\downarrow}$ versus electron energy and external electric field.